\documentstyle[preprint,eqsecnum,tighten,aps]{revtex}
\begin{document}
\draft
\newcommand{\be}{\begin{equation}}
\newcommand{\bestar}{\[\extraspace}
\newcommand{\ee}{\end{equation}}
\newcommand{\eestar}{\]}
\newcommand{\bea}{\begin{eqnarray}}
\newcommand{\eea}{\end{eqnarray}}

\newcommand{\nonu}{\nonumber \\[2mm]}
\newcommand{\ie}{{\it i.e.}\ }

\newcommand{\half}{\frac{1}{2}}
\newcommand{\si}{\sigma}

\newcommand{\cd}{c^{\dagger}}
\newcommand{\Sd}{S^{+}}
\newcommand{\Sz}{S^z}
\newcommand{\etad}{\eta^{\dagger}}
\newcommand{\etaz}{\eta^z}

\newcommand{\up}{\uparrow}
\newcommand{\down}{\downarrow}

\newcommand{\vac}{| \, 0 \, \rangle}
\newcommand{\avac}{\langle \, 0 \, |}
\newcommand{\sumnn}{\sum_{\langle jl \rangle}}
\newcommand{\grst}{|\,\psi_N\,\rangle}

\newcommand{\np}{Nucl.\ Phys.\ }
\newcommand{\phl}{Phys.\ Lett.\ }
\newcommand{\cmp}{Comm.\ Math.\ Phys.\ }
\newcommand{\pr}{Phys.\ Rev.\ }
\newcommand{\phrl}{Phys.\ Rev.\ Lett.\ }
\newcommand{\jop}{J.\ Phys.\ }

\draft

%%%%%%%%%%%%%%% MACROS %%%%%%%%%%%%%%%%%%%%
\newcommand{\sign}{{\rm sign}}
\def\bra#1{\langle #1 |}
\def\ket#1{| #1 \rangle}
\newcommand{\HU}{U\sum_{j=1}^L n_{j\up} n_{j\down}}
\newcommand{\psiNP}{\psi_{P}(N_2)}

%%%%%%%%%%%%%%%%%%%%%%%%%%%%%%%%%%%%%%%%%%%

\preprint{ITP-SB-94-44}

\title{Superconductivity in an exactly solvable
Hubbard model with bond-charge interaction}

\author{Andreas Schadschneider\footnote{Address after Jan.~1, 1995:
Institut f\"ur Theoretische Physik, Universit\"at zu K\"oln,
Z\"ulpicher Strasse 77, D-50937 K\"oln, Germany
(email: {\tt as@thp.uni-koeln.de})}}

\address{Institute for Theoretical Physics\\
State University of New York at Stony Brook\\
Stony Brook, NY 11794-3840, USA\\
email: {\tt aceman@insti.physics.sunysb.edu}}

\date{\today}
\maketitle

\begin{abstract}
The Hubbard model with an additional bond-charge interaction $X$ is
solved exactly in one dimension for the case $t=X$ where $t$ is the
hopping amplitude. In this case the number of doubly occupied sites is
conserved. In the sector with no double occupations the model reduces
to the $U=\infty$ Hubbard model.  In arbitrary dimensions the
qualitative form of the phase diagram is obtained. It is shown that
for moderate Hubbard interactions $U$ the model has superconducting
ground states.
\end{abstract}

\narrowtext
\section{Introduction}
The one-dimensional Hubbard model is since long the prototype of an
exactly solvable model for correlated electrons
\cite{LIEB,KEBOOK}. However, it is difficult to include additional
interactions so that the resulting model is still integrable. In this
paper we present a generalized Hubbard model which apart from the
Coulomb interaction $U$ also contains a bond-charge interaction term
$X$. This model has been studied extensively by Hirsch
\cite{HIRSCH,MARSIG} who argued that it might be relevant for the
description of high-$T_c$ superconductors. For certain values of $X$
and large densities of electrons (small doping) the bond-charge
interaction can lead to an attractive effective interaction between
the holes and the formation of Cooper-pairs. Although this picture was
confirmed by a BCS-type mean-field theory it is desirable to find
exact results confirming this behaviour. A first step in this
direction was made in \cite{BKSZ} where a simplified version of
Hirsch's Hamiltonian has been studied in one dimension by Bethe-Ansatz
methods. Indeed, it has been found that this model has a strong
tendency towards superconductivity (see also \cite{JAP}).

On the other hand, there has recently been a lot of activity in the
investigation of the so-called $\eta$-pairing mechanism of
superconductivity.  This idea has been introduced by Yang \cite{YANGZ}
for the Hubbard model.  It allows for the construction of states
exhibiting off-diagonal-long-range-order (ODLRO). As shown in
\cite{YANGO,SEWELL,NIEH} ODLRO implies also the Meissner effect and
flux quantisation and can thus be regarded as definition of
superconductivity. In the case of the Hubbard model it was found that
the ground state is not of the $\eta$-pairing type \cite{YANGZ}.  The
first model with a truly superconducting ground state of the
$\eta$-pairing type is the supersymmetric Hubbard model introduced by
E\ss ler, Korepin and Schoutens (EKS model) \cite{EKSa,EKSb}. The EKS
model is a Hubbard model which contains additional nearest-neighbour
interactions.  It has some attractive features. Apart from being
exactly solvable by Bethe Ansatz in one dimension it is possible to
determine the $T=0$ phase diagram in arbitrary dimensions. One always
finds a superconducting ground state if $U<U_c$ where the critical
value $U_c$ is positive corresponding to a repulsive on-site Coulomb
interaction.

In \cite{dBKS} it has been shown that an $\eta$-pairing ground state
is not an exotic phenomenon restricted to one special model but may be
found for a large class of Hamiltonians. As a special case the Hubbard
model with bond-charge interaction $X$ is in this class, provided that
$X=t$ where $t$ is the hopping matrix element. It is this model which
will be studied in this paper\footnote{A brief account of some of the
results presented here can be found in \cite{dBKS}.}.

In Sec.~II the symmetries of the Hamiltonian are investigated. It
turns out that the model under investigation is a generalization of
the $U=\infty$ Hubbard model to a model with a conserved (but in
general non-zero) number of doubly occupied sites. Since the
$U=\infty$ Hubbard model has attracted a lot of interest in recent
years \cite{CASPERS,DOUCOT,KOTRLA,OGATA,AUDIT,KRIV,MIELKE,DIAS,KUSM}
it seems worthwhile to study any generalization of it.

In Sec.~III we present the exact solution in one dimension. It will be
shown that the Hamiltonian can be mapped in certain subspaces onto a
spinless free fermion model with twisted boundary conditions where the
twist depends on the subspace considered. Using this fact it is easy
to obtained the spectrum and the zero temperature phase diagram of the
Hamiltonian. It is found that for $U$ not too large the model contains
superconducting ground states of the $\eta$-pairing type. For certain
densities this is true even for moderately repulsive Coulomb
interactions.

In Sec.~IV we construct the qualitative form of the phase diagram in
arbitrary dimensions. Basically it has the same form as the phase
diagram in one dimension although the exact location of all phase
boundaries can not be determined.

Sec.~V contains a summary of the results and a discussion of the
stability of the superconducting ground states.

In the Appendix details of the exact solution in one dimension are
given.

\section{The model and its symmetries}
The Hamiltonian of the Hubbard model with an additional bond-charge
interaction $X$ (or correlated hopping) on a $d$ dimensional lattice
with $L$ sites and periodic boundary conditions is given by
\bea
{\cal H}(X,U)&=&-t\sumnn\sum_{\sigma=\up,\down}
\left(\cd_{j\sigma} c_{l\sigma}+ \cd_{l\sigma} c_{j\sigma}\right)
+\HU \nonu
&&\quad +X\sumnn\sum_{\sigma=\up,\down}
\left(\cd_{j\sigma} c_{l\sigma}+ \cd_{l\sigma} c_{j\sigma}\right)
\left( n_{j,-\sigma}+n_{l,-\sigma}\right)\ .
\label{Ham}
\eea
Here $c_{j\sigma}$, $c_{j\sigma}^\dagger$ are the usual Fermi
operators and $n_{j\sigma}=c_{j\sigma}^\dagger c_{j\sigma}$ is the
corresponding number operator. $\langle jl\rangle$ denotes
nearest-neighbour sites on the $d$-dimensional lattice.  In the
following we will be interested in the special case ${\cal H}(U)=
{\cal H}(X=t,U)$. In this case it is convenient to rewrite the
Hamiltonian in terms of Hubbard operators $X^{ab}=\ket{a}\bra{b}$
($a,b=0,\pm 1, 2$) where $0$ denotes an empty site, $\pm 1$ a site
occupied by a single electron with spin $\up$ or $\down$, and 2 a
doubly occupied site. The fact that for each site $j$ the four
states $\ket{0}$, $\ket{-1}$, $\ket{1}$, and $\ket{2}$ form a basis
of the local Hilbert space leads to the local constraint
$X_j^{00}+X_j^{22}+\sum_{\sigma} X_j^{\sigma \sigma} =1$.

The Hubbard operators obey the following graded commutation rules
\be
[X_j^{ab},X_l^{cd}]_{\pm}=\left(X_j^{ad}\delta_{bc}\pm
X_j^{cb}\delta_{ad} \right)\delta_{jl}\ ,
\label{comm}
\ee
where the anticommutator has to be taken only if both operators are
fermionic, i.e.~change the particle number by one (e.g. $X_j^{\sigma
0}$ or $X_j^{2\sigma}$). The standard fermion operators $c_{j\sigma}$
can be expressed through the Hubbard operators
\be
 c_{j\sigma}^\dagger=X_j^{\sigma 0}+\sigma X_j^{2,-\sigma },\quad
c_{j\sigma}=X_j^{0\sigma}+\sigma X_j^{-\sigma 2}, \quad
n_{j\sigma}=X_j^{\sigma\sigma}+X_j^{22}
\label{cX}
\ee
and vice versa, e.g.
\bea
& &X_j^{\sigma 0}=(1-n_{j,-\sigma})c_{j\sigma}^\dagger,\quad
X_j^{2,-\sigma}=n_{j,-\sigma}c_{j\sigma}^\dagger, \nonu
& &X_j^{\sigma \sigma}=(1-n_{j,-\sigma})n_{j\sigma},\quad
X_j^{00}=(1-n_{j\downarrow})(1-n_{j\uparrow}).
\label{Xc}
\eea

In terms of the Hubbard operators the Hamiltonian (\ref{Ham}) takes
the following form:
\bea
{\cal H}(U)&=&{\cal H}_0+{\cal H}_U \nonu
&=& -t\sumnn\sum_{\sigma=\pm 1} \left( X_j^{\sigma 0} X_{l}^{0\sigma}
  + X_{l}^{\sigma 0} X_j^{0\sigma } + X_j^{\sigma 2} X_{l}^{2\sigma }
  + X_{l}^{\sigma 2} X_j^{2\sigma }\right)
+U\sum_{j=1}^L X_j^{22}\ .
\label{HamX}
\eea

The Hamiltonian ${\cal H}(U)$ has a lot of symmetries, since only two
types of processes are allowed: $(i)$ a particle from a singly
occupied site may hop to an unoccupied neighbour site, or $(ii)$ a
particle with spin $\sigma$ from a doubly occupied site may hop to a
singly occupied (particle with spin $-\sigma$) neighbour site.
Therefore it is clear that ${\cal H}(U)$ conserves not only the total
number $N$ of electrons but also the number $N^{(\sigma)}_1$ of {\it
single} electrons with spin $\sigma$ and the number
$N_2=\sum_{j=1}^L n_{j\up}n_{j\down}$ of doubly occupied sites, i.e.\
$[{\cal H}_0,{\cal H}_U]=0$.  In addition, ${\cal H}_0$ has two
SU(2)-symmetries: apart from the SU(2) spin symmetry, i.e. ${\cal
H}_0$ commutes with ${\cal S}^+ =\sum_{j=1}^L \cd_{j\up} c_{j,\down}$,
${\cal S}^-=\left({\cal S}^+ \right)^\dagger$, ${\cal S}^z={1 \over
2}(N^{(\up)}_1-N_1^{(\down)})$, ${\cal H}_0$ also commutes with the
following so-called $\eta$-operators:
\be
\eta=\sum_{j=1}^Lc_{j\up}c_{j\down},\qquad
\eta^\dagger=\sum_{j=1}^L\cd_{j\down}\cd_{j\up},\qquad
\eta^z={1 \over 2}\left( N - L \right)
\label{eta}
\ee
where $N=N_1 + 2N_2$ (with $N_1=N_1^{(\up)}+N_1^{(\down)}$) is the
total number of particles. In terms of the Hubbard operators
the $\eta$-operators are given by $\eta = -\sum_{j=1}^L X_j^{02}$ and
$\eta^\dagger = -\sum_{j=1}^L X_j^{20}$.

$\eta^\dagger$ creates a double occupation with momentum zero as can
be seen from its form in momentum space $\eta^\dagger = \sum_k
\cd_{k\down}\cd_{-k\up}$ where $\cd_{k\sigma} ={1 \over \sqrt{L}}
\sum_k e^{-ijk}\cd_{j\sigma}$. Thus $\eta^\dagger$ is just the
conventional s-wave pairing operator. Note that in the case of the
Hubbard model the corresponding $\eta$-symmetry is generated by
$\eta_\pi=\sum_{j=1}^L (-1)^j c_{j\up}c_{j\down}$ corresponding to
pairs with momentum $\pi$ \cite{YANGZ}.

The Hamiltonian ${\cal H}(U=0)$ without the Coulomb interaction is
also invariant under a particle-hole transformation:
\bea
{\cal U}c_{j\sigma}{\cal U}^{-1}&=& c_{j\sigma}^\dagger,\qquad
{\cal U}c_{j\sigma}^\dagger{\cal U}^{-1}=c_{j\sigma},\label{ph1}\\
{\cal U}{\cal H}_0 {\cal U}^{-1} &=&{\cal H}_0\ ,
\label{phH0}
\eea
whereas the Coulomb interaction ${\cal H}_U=\HU$ transforms as
\be
{\cal U H}_U {\cal U}^{-1} ={\cal H}_U+U(L-N).
\label{phHU}
\ee
Due to this particle-hole symmetry we can restrict our investigation
to the case $N\leq L$.

In the following especially the conservation of $N_2$ will be
important. It allows to diagonalize the Hamiltonian in subspaces with
fixed $N_2$. For $N_2=0$ the Hamiltonian ${\cal H}_0$ reduces to the
$U=\infty$ Hubbard model
\cite{CASPERS,DOUCOT,KOTRLA,OGATA,AUDIT,KRIV,MIELKE,DIAS,KUSM}
\bea
{\cal H}_\infty&=&-t~\sumnn\sum_{\sigma=\up,\down}
\left( X_j^{\sigma 0} X_{l}^{0\sigma} + X_{l}^{\sigma 0}
 X_j^{0\sigma }\right) \nonu
&=&-t~{\cal P}\sumnn\sum_{\sigma=\up,\down}\left(c_{l\sigma
}^\dagger c_{j\sigma}+ c_{j\sigma}^\dagger c_{l\sigma}\right){\cal P}
\label{Hinfty}
\eea
where ${\cal P}$ is the projector onto the subspace with no doubly
occupied sites. Thus the Hamiltonian (\ref{HamX}) is a generalization
of the the $U=\infty$ Hubbard model to a model with a conserved (but
in general nonzero) number of doubly occupied sites. This is similar
to the EKS-model \cite{EKSa,EKSb} which is a generalization of the
supersymmetric $t$-$J$ model \cite{BARES} to a model with a conserved
number of doubly occupied sites.

\section{Exact solution in one dimension}

In the following the Hamiltonian ${\cal H}_0$ (see (\ref{HamX})) is
diagonalized in one dimension with periodic boundary
conditions\footnote{A solution of an equivalent model with open
boundary conditions has been given in \cite{KLEIN} (see also
\cite{ALIG}).}.

In the subspace $H_{N_1,N_2}$ with $N_1$ singly occupied sites and
$N_2$ doubly occupied sites, i.e.\ $N=N_1+2N_2$ particles,
we define the following basis vectors:
\be
|\vec x,\vec \sigma,\vec b\rangle =
\prod_{\alpha=1}^{N_1} X_{x_\alpha}^{\sigma_\alpha,0}
\prod_{\beta=1}^{L-N_1}X_{y_\beta}^{b_\beta,0}|0\rangle
\qquad (1\leq x_1 < \cdots < x_{N_1} \leq L).
\label{basis}
\ee
$|\vec x,\vec \sigma,\vec b\rangle$ is a state in which the sites
$x_\alpha$ are occupied by a single electron with spin
$\sigma_\alpha$. $\{ y_1,\ldots y_{L-N_1}\} = \{ 1,\ldots, L
\}\setminus \{ x_1,\ldots, x_{N_1} \}$ are the sites occupied by a
boson $b_\alpha$ where $b_\alpha = 2$ \ for a doubly occupied site and
$b_\alpha = 0$ for an empty site. Note that due to (\ref{comm}) all
$X_{y_\beta}^{b_\beta,0}$ commute mutually and with all the
$X_{x_\alpha}^{\sigma_\alpha,0}$.

Next we define the operators ${\cal C}$ and ${\cal C}'$ generating
cyclic permutations of the spins and bosons, respectively:
\bea
{\cal C}\{\sigma_1,\ldots ,\sigma_{N_1}\} &=& \{\sigma_2,\ldots ,
\sigma_{N_1},\sigma_1 \},\label{perm1}\\
{\cal C}'\{b_1,\ldots ,b_{N-N_1}\} &=& \{b_{N-N_1-1},b_1,b_2,\ldots
,b_{N-N_1-1} \}.
\label{perm2}
\eea
Using these operators we can define subspaces of $H_{N_1,N_2}$. For
every spin configuration $\{\sigma_1,\ldots ,\sigma_{N_1}\}$ there
exists a minimal integer $K\geq 1$ such that
\be {\cal C}^K\{\sigma_1,\ldots ,\sigma_{N_1}\} = \{\sigma_1,\ldots
,\sigma_{N_1}\}.
\label{Kdef}
\ee
Clearly for a fully polarized ferromagnetic state we have $K=1$ and
for a N\'eel state $K=2$. In the same way we define the integer $K'$
for the distribution of bosons.

All the states $|\vec x,\vec \sigma,\vec b\rangle$ with $\vec\sigma$
and $\vec b$ characterized by the same integers $K$ and $K'$ span a
subspace $H_{N_1,N_2}(K,K')$ of $H_{N_1,N_2}$. It is now important to
notice that the subspaces $H_{N_1,N_2}(K,K')$ are invariant under the
action of the Hamiltonian ${\cal H}$. This fact is well-known
\cite{CASPERS,KOTRLA} for the $U=\infty$ Hubbard model, i.e.\ $N_2=0$.
Since the local Hamiltonian $h_{j,j+1}$ only permutes bosons with
fermions a spin $\sigma_{\alpha+1}$ will under the action of ${\cal
H}$ always stay ``to the right'' of a spin $\sigma_\alpha$ except for
the case where it moves ``over the boundary'' ($L\to L+1 \equiv 1$).

We can now restrict ourselves to to the diagonalization of the
Hamiltonian in the subspaces $H_{N_1,N_2}(K,K')$. In the appendix
it is shown that in each of these subspaces ${\cal H}$ is equivalent to
a free fermion Hamiltonian $\tilde{\cal H}$ with twisted boundary
conditions:
\be
{\tilde{\cal H}}=-\sum_{j=1}^{L-1}\left(a_j^\dagger a_{j+1}
+a_{j+1}^\dagger a_{j}\right) -\left(e^{iL\Delta\phi}a_L^\dagger a_1
+ e^{-iL\Delta\phi}a_1^\dagger a_L \right)
\label{Htwist}
\ee
which now acts on the ``stripped states'' \cite{CASPERS}
\be
|\vec x\rangle = \prod_{\alpha=1}^Na_{x_\alpha}^\dagger |0\rangle \ .
\label{stripped}
\ee
The $a_j$ are (spinless) fermion operators and the allowed values
$\Delta \phi$ for the twist in the boundary conditions is different
for each subspace:
\bea
\Delta \phi &=&{k'-k \over L}\ , \label{twist}\\
k&=&{2\pi\over K}\nu\quad\qquad (\nu=0,1,\ldots, K-1),\label{kdef}\\
k'&=&{2\pi\over K'}\nu'\quad\qquad (\nu'=0,1,\ldots, K'-1).
\label{kqdef}
\eea
After the canonical transformation $a_j^\dagger \to e^{ij\Delta \phi}
a_j^\dagger$, $a_j \to e^{-ij\Delta \phi}a_j$ the Hamiltonian
(\ref{Htwist}) is seen to be equivalent to the following
translational-invariant free fermion Hamiltonian
\be
{\cal{H}}_{{\rm eff}}=-\sum_{j=1}^L\left(e^{i\Delta\phi}a_j^\dagger
a_{j+1}+ e^{-i\Delta\phi}a_{j+1}^\dagger a_{j}\right)
\label{Heff}
\ee
from which the eigenvalues of $\cal H$ can be obtained easily by
Fourier transformation:
\be
E=-2\sum_{\nu}\cos(q_\nu+\Delta\phi)n_{q_\nu}
\label{energy}
\ee
with $\sum_\nu n_{q_\nu}=N_1$ and $n_{q_\nu}=1~(0)$ if the mode
$q_\nu$ is occupied (not occupied). The wavenumbers $q_\nu$ can take
the values $q_\nu={2\pi\over L}\nu$ ($\nu=-{L\over 2}+1,
\ldots,{L\over 2}$).

In the ground state the $q_\nu$ are as symmetric as possible around
$\nu=0$.  For $N_1$ even we have $n_{q_\nu}=1$ for $\nu=-{N_1\over
2}+1,\ldots, {N_1\over 2}$ and thus
\be
E=-2{\cos\left({\pi\over L}+\Delta\phi\right)\over
\sin\left(\pi/L\right)} \sin(N_1\pi/L)\ .
\label{grenereven}
\ee
For $N_1$ odd we have to choose $\nu=-{N_1-1\over 2}+1,\ldots,
{N_1-1\over 2}$ yielding
\be
E=-2{\cos\left(\Delta\phi\right)\over \sin\left(\pi/L\right)}
\sin(N_1\pi/L).
\label{grenerodd}
\ee
This shows that in the ground state one has
\be
\Delta\phi=\cases{-\pi/L &($N_1$ even),\cr
0\quad &($N_1$ odd),\cr}
\label{grtwist}
\ee
which gives the ground state energy
\be
E_0=-2{\sin(N_1\pi/L)\over \sin\left(\pi/L\right)}.
\label{E0}
\ee
The ground state energy only depends on the number $N_1$ of
singly occupied sites and is independent of the number $N_2$ of doubly
occupied sites. Therefore the ground state energy of (\ref{HamX}) is
$E_0(N_1,N_2,U)=E_0+UN_2$.

The mapping of the original model onto {\it spinless} free fermions
(\ref{Heff}) implies that the eigenvalues of (\ref{HamX}) are highly
degenerate in general.

\section{Phase diagram at T=0}

In this section we determine the phase diagram of the Hamiltonian
(\ref{HamX}) in the $U$-$D$--plane where $D=N/L$ is the particle
density. Firstly, we consider the one-dimensional case which can be
treated exactly using the results of the previous section. Then we
show that the phase diagram in higher dimensions has qualitatively the
same form as in one dimension.  We find the same phases although we
cannot determine all phase boundaries exactly. Finally we calculate
correlation functions for generalized $\eta$-pairing states which
appear as ground states in certain parameter regions.

\subsection{Phase diagram in one dimension}
In order to determine the phase diagram in the $U$-$D$--plane for the
one-dimensional case we have to minimize the ground state
energy\footnote{From now on we work in the thermodynamic limit
$L,N_\alpha \to\infty$ with $D_\alpha = N_\alpha/L$ fixed ($\alpha =
1,2$).}  $E_0(D_1,D_2,U)/L=-{2\over \pi}\sin(D_1\pi)+UD_2$ for a fixed
particle density $D=D_1+2D_2$.  A simple calculation yields
\be
D_1=\cases{0 &$(U\leq -4)$,\cr
{1\over\pi}\arccos(-U/4) &$(-4\leq U\leq U_c)$,\cr
D &$(U\geq U_c)$,\cr}
\label{D1}
\ee
where $U_c(D)=-4\cos(\pi D)$.

Due to the particle-hole symmetry it is sufficient to discuss the phase
diagram for $D\leq 1$. We find four different phases which will be
discussed separately in the following.
\\
\underbar{Regime I}: $\quad U\leq -4$\\
{}From (\ref{D1}) we see the ground state contains only doubly-occupied
sites and no single electrons. In this case the ground state energy is
simply $E_0(N,0,U)/L=UN/2$. In the absence of single electrons the
double occupations are static and all states with the same number
$N_2=N/2$ of doubly-occupied sites have the same energy. Among these
states are the generalized $\eta$-pairing states
\bea
\ket{\psi_P} &=&\bigl(\eta_P^\dagger\bigr)^{N/2}\ket{0}, \label{psiP}\\
\eta_P^\dagger &=& \sum_{j=1}^L e^{iPj}c_{j\down}^\dagger
c_{j\up}^\dagger
\label{etaP}
\eea
with momentum $P={2\pi\over L}\nu$ ($\nu=0,1,\ldots,L-1$).  These states
are ground states for all chain lengths $L$, not only in
the thermodynamic limit.  All of these states show ODLRO,
i.e.\footnote{See also Eq.\ (\ref{etacorr}) below.}
\be
\lim_{|l-j|\to\infty} {\langle\psi_P|c_{j\down}^\dagger c_{j\up}^\dagger
c_{l\up} c_{l\down}|\psi_P\rangle \over \langle\psi_P| \psi_P\rangle}
\neq 0 \ .
\label{ODLRO}
\ee
As a consequence these ground states are superconducting since it has
been shown that ODLRO also implies the Meissner effect and flux
quantisation \cite{YANGO,SEWELL,NIEH}.
\\
\underbar{Regime II}: $\quad -4<U<U_c(D)=-4\cos(\pi D)$\\
For $-4< U < U_c(D)$ the ground state has both a finite density of
single electrons and of double occupations. Again it is highly degenerate.
Here the ground state energy is $E_0(D_1,D_2,U)/L = -{1\over
2\pi}\sqrt{16-U^2} + U D_2$. Among these
ground states are the $\eta$-states $\bigl(\eta_0^\dagger\bigr)^{N_2}
\ket{U=\infty}$ where $\ket{U=\infty}$ stands for an arbitrary ground
state of the $U=\infty$ Hubbard model at particle density $D_1={1\over\pi}
\arccos(-U/4)$ and $N_2$ is then obtained from $D_2=N_2/L =(D-D_1)/2$.
Again these states have ODLRO (see the discussion of correlation functions
below) and are thus superconducting. It is interesting that this
superconducting phase extends into the region of positive $U$, i.e.\
Coulomb repulsion.
\\
\underbar{Regime III}: $\quad U\geq U_c(D)=-4\cos(\pi D)$ and $D<1$\\
In regime III we have no doubly occupied sites in the ground state and
every ground state of the $U=\infty$ Hubbard model is therefore also a
ground state of (4). Regime III' with $D>1$ is obtained by using the
particle-hole symmetry (\ref{ph1})-(\ref{phHU}).
\\
\underbar{Regime IV}: $\quad U\geq U_c(D)=-4\cos(\pi D)$ and $D=1$\\
This case is much like regime III, but now the ground states
$\ket{U=\infty}$ are insulating. The point $(D=1,U=4)$ corresponds to a
metal-insulator transition.
\\
The complete phase diagram for the one-dimensional case is shown in
Fig.~1 (see also \cite{dBKS,ALIG}).

\subsection{Phase diagram in arbitrary dimensions}
In higher dimensions the phase diagram can be constructed along the
lines of \cite{EKSb}. In \cite{dBKS} it has been shown that
$\ket{\psi_0}=\bigl(\eta_0^\dagger\bigr)^{N/2}\ket{0}$ will be a
ground state for $U\leq -2Z$ where $Z$ is the number of nearest
neighbour sites in the $d$-dimensional lattice. In order to construct
the full phase diagram we need the following three properties:
\begin{description}
\item[$1)$] $\eta$-symmetry, i.e.\ $[{\cal H}_0,\eta_0 ]=0$,
\item[$2)$] conservation of the number $N_2$ of doubly occupied sites,
\item[$3)$] for $U=0$ the ground state energy does not depend on the
number $N_2$ of doubly occupied sites.
\end{description}
The first two properties have already been demonstrated in Sec.~II,
property 3) can be proven by generalizing the argumentation of \cite{SUTH}
where the analogous property for the EKS-model has been derived.

\begin{figure}
\caption{$T=0$ phase diagram in one dimension. The phase diagram in
higher dimensions has qualitatively the same form.}
\label{FigPhase}
\includegraphics{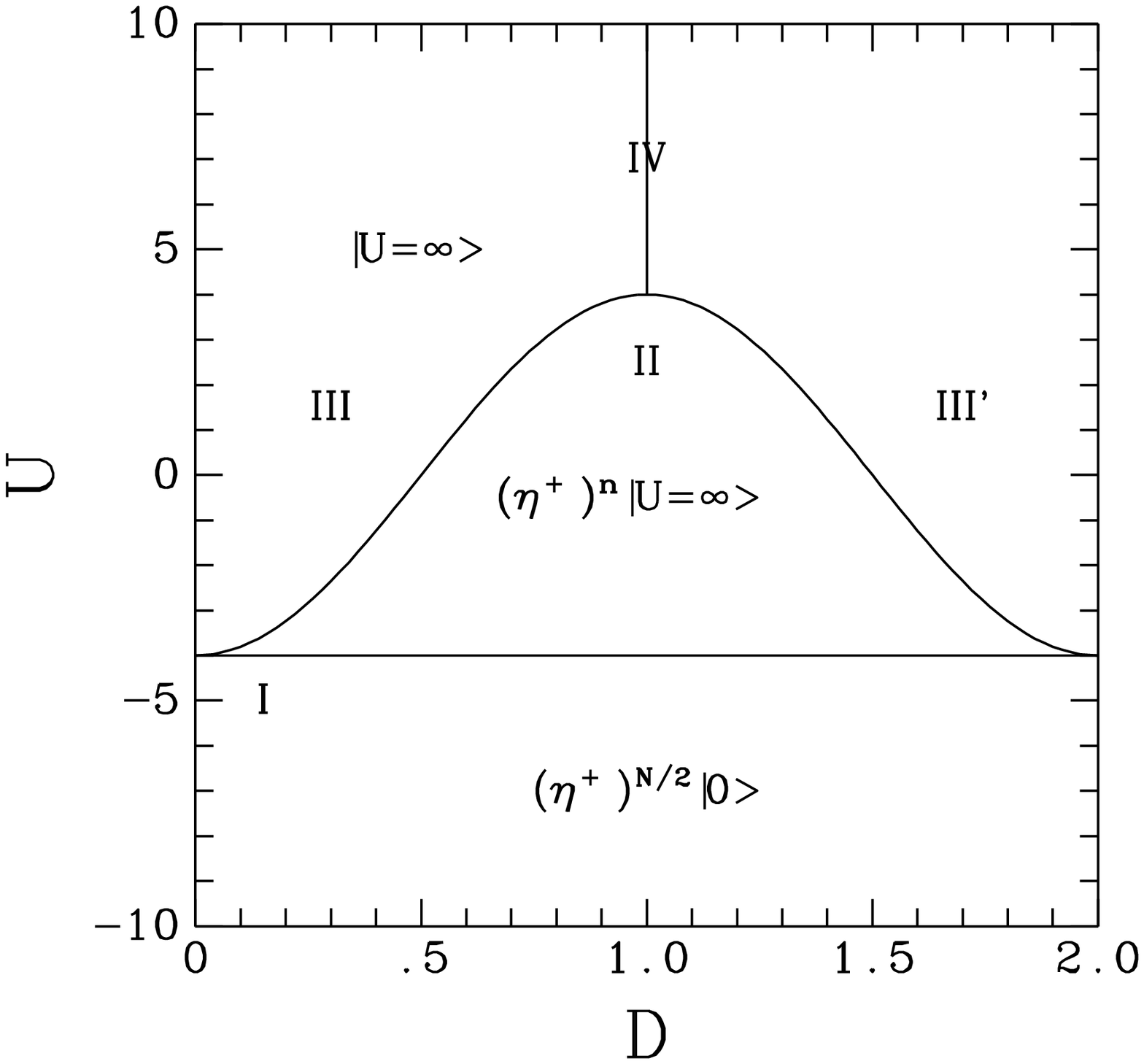}
\end{figure}

Using 1)--3) the qualitative form of the phase diagram can be
established in complete analogy with \cite{EKSb} by first considering
the grand canonical ensemble and then translating the results into the
canonical ensemble. One finds a phase diagram which looks very similar
to that of the one-dimensional case (Fig.~\ref{FigPhase}). The same
phases appear, only the location of the phase boundaries
changes. Except for the boundary between regimes I and II (see
\cite{dBKS}) we have not been able to determine them exactly. An
interesting open question is thus whether the superconducting regime
II extends into the positive-$U$ region in all dimensions.

\subsection{Correlation functions}
Correlation functions with respect to the $\eta$-pairing states can be
calculated in a straightforward way \cite{EK}. Let $\ket{\phi}$ be a
state with $\eta_P \ket{\phi} = 0$ and $\eta^z \ket{\phi} = {1 \over
2} (N_1-L) \ket{\phi}$, e.g.\ $\ket{\phi}$ is a state with $N_1$
singly and no doubly occupied sites. In our case we have
$\ket{\phi}=\ket{0}$ or $\ket{\phi} =\ket{U=\infty}$. For the
$\eta$-pairing states $\ket{\psiNP} =
\bigl(\eta_P^\dagger\bigr)^{N_2}\ket{\phi}$ we denote the expectation
value of an arbitrary operator ${\cal O}$ by $\langle {\cal O}\rangle
= {\bra{\psiNP} {\cal O} \ket{\psiNP} \over \bra{\psiNP}
\psiNP\rangle}$. We now can express any correlation function with
respect to $\ket{\psiNP}$ through correlation functions with respect to
$\ket{\phi}$. With $\langle {\cal O} \rangle_\phi = {\bra{\phi}{\cal
O}\ket{\phi} \over \bra{\phi}\phi\rangle}$ we find (for $L\to\infty$,
$D_\alpha=N_\alpha/L$ and $j\neq l$):
\bea
\langle \eta_j^\dagger\eta_l\rangle &=& \langle c_{j\down}^\dagger
  c_{j\up}^\dagger c_{l\up} c_{l\down} \rangle
  =e^{iP(l-j)}\, {(1-D_1-D_2)D_2\over (1-D_1)^2}\, \langle (1-n_j)(1-n_l)
  \rangle_\phi\ , \label{etacorr}\\
\langle c_{j\sigma}^\dagger c_{l\tau} \rangle &=& {1-D_1-D_2 \over 1-D_1}
  \, \langle c_{j\sigma}^\dagger c_{l\tau} \rangle_\phi -\sigma\tau
  e^{iP(l-j)}{D_2 \over 1-D_1}\,  \langle c_{l,-\tau}^\dagger
  c_{j,-\sigma} \rangle_\phi \ , \label{ccorr}\\
\langle n_{j\sigma}n_{l\tau}\rangle &=&
  \langle n_{j\sigma}n_{l\tau}\rangle_\phi
  +{D_2\over 1-D_1}\Bigl( \langle n_{j\sigma}(1-n_l)\rangle_\phi +
  \langle n_{l\tau}(1-n_j)\rangle_\phi\Bigr)\nonu
& &\qquad +\left({D_2\over 1-D_1}\right)^2
  \langle (1-n_{l})(1-n_j)\rangle_\phi\ , \label{ncorr}\\
\langle S_j^\alpha S_l^\alpha\rangle &=& \langle S_j^\alpha S_l^\alpha
  \rangle_\phi \qquad (\alpha=x,y,z).
\label{Scorr}
\eea
These results hold in all dimensions in the thermodynamic limit.
Correlators for phase I (where $\ket{\phi}=\ket{0}$) are trivial.  The
asymptotics for the correlation functions in phase II (where
$\ket{\phi} =\ket{U=\infty}$) in one dimension can be obtained using
the results of \cite{FRAHM,KAWA,PARSOR,SHIBA}.  The result for the
$\eta$-correlator $\langle \eta_j^\dagger\eta_l\rangle$ proves the
existence of ODLRO since the value of the limit in (\ref{ODLRO}) is
found to be $e^{iP(l-j)}(1-D_1-D_2)D_2 \neq 0$.

\section{Conclusions}
In this paper a Hubbard model with an additional bond-charge
interaction $X$ has been investigated at the special point $X=t$ where
the number of doubly occupied sites is conserved. In one dimension the
complete spectrum of the Hamiltonian could be determined exactly by
mapping onto a system of spinless free fermions with twisted boundary
conditions.  In arbitrary dimensions the $T=0$ phase diagram in the
$D$-$U$-plane has been obtained. One finds four phases, two of them
containing ground states of the $\eta$-pairing type. These states
exhibit ODLRO and thus are superconducting.  Therefore the model
(\ref{Ham}) provides a new example of a purely electronic model with a
superconducting phase. The Cooper pairs have zero-size which might be
regarded as an approximation to the small coherence length found
experimentally in high-$T_c$ superconductors\footnote{For reviews on
other models of local-pair superconductors, see e.g.\
\cite{MICNAS,JONGH} and references therein.}.

Let us finally discuss the question of stability of the
superconducting phases. As in the EKS model all interaction constants
(apart from the Coulomb interaction $U$) have to take certain values
in order to allow for the $\eta$-pairing ground states. Especially the
bond-charge interaction $X$ has to be equal to the hopping matrix
element $t$ in order to guarantee the conservation of local pairs. One
might wonder if the superconducting properties will survive if one
allows for a decay of Cooper pairs (i.e.~$X\neq t$). The results
obtained in \cite{LEE} suggest that this is indeed the case.

For the EKS model other perturbations of coupling constants have been
investigated \cite{AKS}. It has been found that in one dimension these
perturbations destroy superconductivity, but in higher dimensions the
supercondcuting phase is likely to be stable under such perturbations.

A similar analysis is also desirable for the model presented here. It
would be interesting to find a perturbation by allowing for (small)
additional nearest-neighbour interactions which lifts the large ground
state degeneracy in such a way that only (some) of the $\eta$-pairing
states remain as ground states. In \cite{STRACK,OVCH} the ground state
for the model (\ref{Ham}) has been obtained in the presence of an
additional nearest-neighbour Coloumb interaction $V$ at half-filling
in arbitrary dimensions. In that case the ground state is unique
(apart from a trivial twofold degeneracy) for $U < 2ZV-Z\max(2t,V)$,
but not superconducting.

These questions are currently under investigation using perturbation theory
and exact diagonalizations of small systems. Results will be presented
in a future publication.

\acknowledgments

The author likes to thank Vladimir Korepin and Jan de Boer for
useful discussions and the Deutsche Forschungsgemeinschaft
for financial support.

\appendix
\section*{}
\setcounter{equation}{0}
In this appendix we extend the method used in \cite{CASPERS} to map the
$U=\infty$ Hubbard model onto a free fermion Hamiltonian with twisted boundary
conditions to our case.

First we investigate the action of the local Hamiltonian (we set $t=1$
from now on)
\bea
h_{j,j+1} &=& h_{j,j+1}^{(+)}+h_{j,j+1}^{(-)}, \qquad
h_{j,j+1}^{(-)} = \left( h_{j,j+1}^{(+)} \right)^\dagger\ , \nonu
h_{j,j+1}^{(+)}&=&-\sum_{\sigma = \up,\down}\sum_{\delta = 0,2}
X_j^{\sigma\delta}X_{j+1}^{\delta\sigma}
\label{A1}
\eea
on the states $|\vec x,\vec\sigma,\vec b\rangle$. We restrict ourselves to
the investigation of $h_{j,j+1}^{(+)}$ since $h_{j,j+1}^{(-)}$ can be
treated analogously.

It is easy to see that $h_{j,j+1}^{(+)} |\vec x,\vec\sigma,\vec b\rangle
=0$ if either $j\in \{x_1,\ldots,x_{N_1}\}$ or $j+1 \notin \{x_1,\ldots,
x_{N_1}\}$. Therefore it is sufficient to consider the case $j+1 = x_\gamma
\in \{x_\alpha\}$, $j=y_{\gamma'}\in\{ y_\beta\}$. A straightforward
calculation yields for $1\leq j\leq L-1$
\be
h_{j,j+1}^{(+)} |\vec x,\vec\sigma,\vec b\rangle = -|\vec{x'},
\vec\sigma,\vec b\rangle
\label{A2a}
\ee
with
\be
x'_\alpha = \cases{x_\alpha &\qquad$(\alpha \neq \gamma)$\cr
j &\qquad$(\alpha = \gamma)$\cr}
\label{A2}
\ee
and for $j=L$
\be
h_{L,1}^{(+)} |\vec x,\vec\sigma,\vec b\rangle = (-1)^{N_1}|\vec{x''},
{\cal C}\{\sigma_\alpha\},{\cal C}'\{b_\beta\}\rangle
\label{A3a}
\ee
with
\be
x''_\alpha = \cases{x_{\alpha+1} &\qquad$(\alpha =1,\ldots,N_1-1)$\cr
L &\qquad$(\alpha = N_1)$.\cr}
\label{A3b}
\ee

In the subspace $H_{N_1,N_2}(K,K')$ we introduce the following
states:
\be
|\vec x,k, k'\rangle=\sum_{l=0}^{K-1}\sum_{m=0}^{K'-1}
e^{ikl}e^{-ik'm}|\vec x,{\cal C}^l\{ \sigma_\alpha\},
{\cal C}'^m\{  b_\beta\}\rangle
\label{A4}
\ee
where $k$ and $k'$ are given by (\ref{kdef}) and (\ref{kqdef}), respectively.
$h_{j,j+1}^{(+)}$ acts on these spaces in a simple manner:
\be
h_{j,j+1}^{(+)} |\vec x,k,k'\rangle =
\cases{-|\vec{x'},k,k'\rangle &$\qquad (j=1,\ldots,L-1)$\cr
(-1)^{N_1}e^{-ik}e^{ik'}|\vec{x''},k,k'\rangle &$\qquad (j=L)$\cr}
\label{A5}
\ee
in the case $j+1\in\{x_\alpha\}$ and $j\in\{y_\beta\}$ and is zero
otherwise. Thus $\sum_{j=1}^L h_{j,j+1}^{(+)}$ acts on the states
$|\vec x,k,k'\rangle$ in the same way as $-\sum_{j=1}^{L-1}
a_j^\dagger a_{j+1} - e^{iL\Delta \phi}a_L^\dagger a_1$ acts on the
``stripped states'' (\ref{stripped}) where the $a_j$, $a_j^\dagger$
are spinless fermion operators and $\Delta\phi$ is given by
(\ref{twist}).

\vfill
\vskip 1truecm

\frenchspacing
\begin{references}

\bibitem{LIEB} E.H.~Lieb, F.Y.~Wu: Phys.\ Rev.\ Lett.\ {\bf 20}, 1445 (1968)

\bibitem{KEBOOK} For a collection of reprints see {\em Exactly Solvable Models
of Strongly Correlated Electrons}, eds.\ V.\ E.\ Korepin and F.\ H.\ L.\
E\ss ler (World Scientific 1994)

\bibitem{HIRSCH}  J.\ E.\ Hirsch: Phys.~Lett.\ {\bf 134A}, 451 (1989); Physica
{\bf 158C}, 326 (1989); Phys.\ Rev.\ {\bf B43}, 11400 (1991)

\bibitem{MARSIG} J.\ E.\ Hirsch, F.~Marsiglio: Phys.\ Rev.\ {\bf B39}, 11515
(1989); {\bf B41}, 2049 (1990);\\
F.~Marsiglio, J.\ E.\ Hirsch: Phys.~Rev.\ {\bf B41}, 6435 (1990)

\bibitem{BKSZ} R.\ Z.\ Bariev, A.\ Kl\"umper, A.\ Schadschneider, J.\ Zittartz:
J.\ Phys.\ {\bf A26}, 1249 and 4863 (1993)

\bibitem{JAP} G.~Japaridze, E.~M\"uller-Hartmann: Ann.~Physik {\bf 3}, 163
(1994)

\bibitem{YANGZ} C.\ N.~Yang: Phys.\ Rev.\ Lett.\ {\bf 63}, 2144 (1989) ; \\
  C.N.~Yang and S.~Zhang, Mod.\ Phys.\ Lett.~{\bf B4}, 759 (1990)

\bibitem{YANGO}  C.\ N.~Yang: Rev.\ Mod.\ Phys.~{\bf 34}, 694 (1962)

\bibitem{SEWELL}  G.\ L.~Sewell: J.\ Stat.\ Phys.~{\bf 61}, 415 (1990)

\bibitem{NIEH} H.\ T.~Nieh, G.~Su, B.\ M.~Zhao: to be published in
Phys.\ Rev.\ {\bf B} (Stony Brook preprint ITP-SB-94-12, cond-mat/9411025)

\bibitem{EKSa}  F.~H.~L.~E\ss ler, V.~E.~Korepin, K.~Schoutens:
Phys.~Rev.~Lett.~{\bf 68}, 2960 (1992)

\bibitem{EKSb}  F.~H.~L.~E\ss ler, V.~E.~Korepin, K.~Schoutens:
Phys.~Rev.~Lett.~{\bf 70}, 73 (1993)

\bibitem{dBKS} J.~de Boer, V.~E.~Korepin, A.~Schadschneider: submitted to
Phys.\ Rev.\ Lett. (Stony Brook preprint ITP-SB-94-18, cond-mat/9405035)

\bibitem{CASPERS} W.~J.~Caspers, P.~L.~Iske: Physica {\bf A157}, 1033 (1989)

\bibitem{DOUCOT} B.~Doucot, X.~G.~Wen: Phys.~Rev.~{\bf B40}, 2719 (1989)

\bibitem{KOTRLA} M.~Kotrla: Phys.~Lett.~{\bf 145A}, 33 (1990)

\bibitem{OGATA} M.~Ogata, H.~Shiba: Phys.~Rev.~{\bf B41}, 2326 (1990)

\bibitem{AUDIT} P.~Audit: J.\ Phys.\ {\bf A23}, L389 (1990);\\
P.~Audit, T.T.~Truong: Phys.\ Lett.\ {\bf A145}, 309 (1990)

\bibitem{KRIV} V.~Y.~Krivnov, A.~A.~Ovchinnikov, V.~O.~Cheranovskii:
Theor.~Math.\ Phys.\ {\bf 82}, 151 (1990)

\bibitem{MIELKE} A.~Mielke: J.~Stat.~Phys.~{\bf 62}, 509 (1991)

\bibitem{DIAS} R.\ G.~Dias, J.M.B.~Lopes dos Santos: J.~Physique {\bf I2},
1889 (1991)

\bibitem{KUSM} F.\ V.~Kusmartsev: J.~Phys.~{\bf CM3}, 3199 (1991)

\bibitem{BARES} P.-A.\ Bares, G.\ Blatter and M.\ Ogata: Phys.~Rev.~{\bf B44},
130 (1991) and references therein

\bibitem{KLEIN} D.~J.~Klein: Phys.~Rev.~{\bf B8}, 3452 (1973)

\bibitem{ALIG} A.~A.~Aligia, L.~Arrachea: Phys.\ Rev.\ Lett.\ {\bf 73}, 2240
(1994)

\bibitem{SUTH} B.~Sutherland: Phys.\ Rev.\ {\bf B12}, 3795 (1975)

\bibitem{EK} V.\ E.\ Korepin, F.~H.~L.~E\ss ler: in {\it Correlation Effects
in Low-Dimensional Electron Systems}, Springer Series in Solid-State Sciences
{\bf 118} (1994), eds.\ A.~Okiji, N.\ Kawakami

\bibitem{FRAHM} H.\ Frahm, V.\ E.\ Korepin: Phys.~Rev.~{\bf B42}, 10553 (1990)

\bibitem{KAWA} N.\ Kawakami, S.-K.\ Yang: Phys.\ Lett.\ {\bf A148}, 359 (1990)

\bibitem{PARSOR} A.\ Parola, S.\ Sorella: Phys.~Rev.~Lett.~{\bf 64}, 1831
(1990); Phys.\ Rev.\ {\bf B45}, 13156 (1992)

\bibitem{SHIBA} H.\ Shiba, M.\ Ogata: Prog.\ Theor.\ Phys.\ Suppl.\ {\bf 108},
265 (1992)

\bibitem{MICNAS} R.~Micnas, J.~Ranninger and S.~Robaszkiewicz: Rev.~Mod.~Phys.\
{\bf 62}, 113 (1990)

\bibitem{JONGH} L.\ J.\ de Jongh: Physica {\bf C161}, 631 (1989)

\bibitem{LEE} R.~Friedberg, T.D.~Lee: Phys.~Rev.~{\bf B40}, 6745 (1989)

\bibitem{AKS} G.\ Albertini, V.\ E.\ Korepin, A.\ Schadschneider: Stony Brook
preprint ITP-SB-94-45 (cond-mat/9411051)

\bibitem{STRACK} R.~Strack, D.~Vollhardt: Phys.~Rev.~Lett.~{\bf 70}, 2637
(1993)

\bibitem{OVCH} A.~A.~Ovchinnikov: Mod.~Phys.~Lett.~{\bf B7}, 1397 (1993)

\end{references}
\end{document}